\documentclass[Journal]{IEEEtran}
\usepackage{cite}
\usepackage{epsfig,graphics,mathrsfs,amssymb,amsmath,bm,color,yfonts,txfonts,multirow,slashbox}
\usepackage{subfigure}
\usepackage{algorithm2e}

\begin{document}

\title{Throughput Enhancement of Multicarrier Cognitive M2M Networks: Universal-Filtered OFDM Systems}
\author{Mirza Golam Kibria, {\it{Member, IEEE}}, Gabriel Porto Villardi, {\it{Senior Member, IEEE}}, Kentaro Ishizu, and Fumihide Kojima, {\it{Member, IEEE}}
 \thanks{The authors are with Wireless Network Research Institute, National Institute of Information and Communications Technology (NICT), Yokosuka Research Park, Japan 239-0847 (e-mails: \{mirza.kibria, gpvillardi, ishidu, f-kojima\}@nict.go.jp).
 
 Copyright (c) 2012 IEEE. Personal use of this material is permitted. However, permission to use this material for any other purposes must be obtained from the IEEE by sending a request to pubs-permissions@ieee.org. }
 }

\maketitle

\begin{abstract}

We consider a cognitive radio network consisting of a primary cellular system and a secondary cognitive machine-to-machine (M2M) system, and study the throughput enhancement problem of the latter system employing universal-filtered orthogonal frequency division multiplexing (UF-OFDM) modulation. The downlink transmission capacity of the cognitive M2M system is thereby maximized, while keeping the interference introduced to the primary users (PUs) below the pre-specified threshold, under total transmit power budget of the secondary base station (SBS). The performance of UF-OFDM based CR system is compared to the performances of OFDM-based and filter bank multicarrier (FBMC)-based CR systems. We also propose a near-optimal resource allocation method separating the subband and power allocation. The solution is less complex compared to optimization of the original combinatorial problem. We present numerical results that show that for  given interference thresholds of the PUs and maximum transmit power limit of the SBS, the UF-OFDM based CR system exhibits intermediary performance in terms of achievable capacity compared to OFDM and FBMC-based CR systems. Interestingly, for a certain degree of robustness of the PUs, the UF-OFDM performs equally well as FBMC. Furthermore, the percentage rate-gain of UF-OFDM based CR system increases by a large amount when UF-OFDM modulation with lower sidelobes ripple is employed. Numerical results also show that the proposed throughput enhancing method despite having lower computational complexity compared to the optimal solution achieves near-optimal performance.

\end{abstract}

\IEEEpeerreviewmaketitle
\section{Introduction}

\IEEEPARstart{S}{tudies} have found that  large portion of the allocated/licensed bands is severely underutilized in both spatial and time domain\cite{FCC,Cabric}. Allocating frequency bands exclusively to specific users/operators provides no guarantee that the bands are used efficiently. One promising solution to this spectrum scarcity problem is to use cognitive radio (CR) technology \cite{Haykin}. In a CR system, users of the primary (licensed) system (PUs) have the exclusive right or highest priority to access the assigned spectrum, but secondary (unlicensed) system can use the licensed spectrum opportunistically as long as it does not cause harmful interference to the primary system. In fact, a major issue in interference-tolerant CR networks in 5th generation (5G) has been how to effectively manage the mutual interference of primary and secondary systems\cite{Wang}.

Machine-to-machine (M2M) communications is an emerging and imminent communication paradigm. It enables ubiquitous and autonomous connectivity between devices and requires minimal or no human intervention \cite{Whitehead,Wu}. M2M technology establishes intelligent communication between things, i.e., it acts as an enabling technology for the realisation and implementation of Internet-of-Things (IoT)\cite{Li}. Furthermore, M2M communications is regarded one potentially disruptive technology that will lead changes in both architectural and component design for 5G networks\cite{Boccardi}. The authors in \cite{Boccardi,Fettweis} argue that 5G networks should have native support for M2M communications.  
Furthermore, 3GPP has been investigating the feasibility of low-cost machine type communications terminal class in long-term evolution (LTE) networks Release 12\cite{Astely}. Some of the main technical challenges of conventional capillary or cellular M2M networks \cite{Aijaz} include spectrum scarcity or resource constraints, interference between co-existing M2M networks due to a multitude of connected devices, energy efficiency and device heterogeneity.

To overcome these technical challenges, the authors in \cite{Zhang1} proposed a new communication paradigm, namely cognitive M2M (M2M communications employing CR technology), which enhances flexibility, efficiency and reliability of M2M communications. Because of its dynamic spectrum access capability, cognitive M2M not only enhances the spectrum utilization efficiency but also exploits alternate spectrum opportunities. Additionally, cognitive M2M is inherently armed to handle the challenges of energy efficiency and interference management. Moreover, cognitive M2M unveils new application areas for M2M communications. Regardless of active research on conventional M2M communications over the last few years, cognitive M2M communications still remains a greatly unexplored area with only a small number of studies. 

In cognitive M2M communications, each terminal has low traffic volume. Furthermore, the low-cost machine type communications devices have limited mobility, and typically operate with low transmission power. In general, the underlying available channel is divided into a number of fixed length time slots, each able to carry a small single packet, and is based on time division duplex (TDD) mode of operation. Additionally, in cognitive M2M bidirectional communication, the system needs to have support for small control message signaling, e.g., uplink sounding and downlink synchronization. In low-latency and delay sensitive M2M application, e.g., smart grid recovery operation, the system should support very fast access to the network. Therefore, the modulation scheme employed in multicarrier cognitive M2M communications needs to (i) offer the ability to enable faster TDD switching, (ii) support transmission with very short transmission time intervals, (iii) support small control message signaling, and (iv) support efficient small packet transmission.

{\bf{\textit{Existing Multicarrier CR Systems and Their Limitations.}}}
Multicarrier modulation techniques have been regarded as strong contenders for CR systems because of their flexibility in allocating radio resources to the users of the secondary system (SUs). Mutual interference between the primary and secondary system is regarded as a performance limiting factor since (i) PU and SU are deployed on adjacent bands and they may employ different access technologies \cite{Farhang}. The transmitting power and the spectral separation between the primary and secondary systems decide the intensity of the mutual interference. The PUs in orthogonal frequency division multiplexing (OFDM) based CR system, experience high interference caused by the secondary system transmission due to the presence of strong sidelobes in its filter frequency response. Furthermore, addition of cyclic prefix (CP) deteriorates spectral efficiency.

Finding an appropriate substitute for OFDM remains a fundamental issue in 5G technology. A basic requirement of 5G is a flexible air interface where the multicarrier attributes like subcarrier spacing is optimisable depending on specific system requirements. 
One of the contender waveforms is filter-bank multicarrier (FBMC)\cite{Schaich,Wunder}. However, For short burst transmissions and control message signaling channels, FBMC suffers from lack of efficiency \cite{Bellanger}. FBMC also exhibits high time domain overheads \cite{Schaich2}, which is not suitable for burst-type (or small packet) data transmission. Although FBMC is better suited than OFDM in theory, practical considerations determinate many issues of the former technology. These drawbacks and limitations exhibited by OFDM and FBMC have led some companies and organisations to transfer their focus to finding a multicarrier modulation scheme suitable for cognitive M2M applications.

{\bf{\textit{Universal-filtered OFDM (UF-OFDM), A Suitable Choice.}}}
Recently, a new multicarrier waveform called universal-filtered OFDM (UF-OFDM)\cite{Schaich,Wunder}, also referred to as UF-OFDM, has attracted a great deal of attention due to its better efficiency in handling IoT traffic than OFDM. The latter is currently the transport mechanism base for LTE and LTE-Advanced systems, and it applies filtering functionality to the  whole frequency band. On the other hand, FBMC applies filtering on a per subcarrier, therefor, requiring very long filter lengths. Multicarrier scheme UF-OFDM filters a group of consecutive subcarriers, therefore, shortening the filter length compared to FBMC, and becomes adjustable depending on types of applications. As a result, UF-OFDM can be a well-suited modulation technique for CR systems including option for short burst transmissions with low power consumption and high efficiency, e.g., cognitive M2M communications, MTC communications and uplink control signaling\cite{Vakilian,Wild}.

UF-OFDM can be regarded as a generalization of this principle that collects the advantages exhibited by OFDM and FBMC while avoiding their drawbacks.  According to \cite{Schaich2}, UF-OFDM offers the ability to enable faster TDD switching: (i) to support transmission with very short transmission time interval, (ii) to support efficient small packet transmission, (iii) to support small control signaling messages, e.g., uplink sounding and downlink synchronization, which make it well-suited for short packet transmission. In addition, UF-OFDM offers great advantages for a network with distributed transmitters and efficiently reduce the intercarrier interference resulting from poor time/frequency synchronization \cite{Andrews}.

Note that UF-OFDM with filter length 1 is identical to non-CP OFDM. Due to this close relationship, reusing existing OFDM transceiver parts is easy. At the UF-OFDM receiver, after 2-$N$ fast Fourier transform, picking each second output and dividing by the frequency response of the filter, yields the same frequency domain scalar per-subcarrier processing as in OFDM. As a result, all existing OFDM channel estimation algorithms can be directly reused by UF-OFDM.
 In addition, FBMC is not orthogonal with respect to the complex gain as it employs offset-QAM\cite{Bellanger1}. As a result, additional guardband needs to be incorporated to separate uplink transmissions (same hold for downlink transmission if complex precoding is applied) from different M2M entities. On the other hand, UF-OFDM is orthogonal with respect to the complex plain as it employs QAM like OFDM. Therefore, it is not necessary to add extra guardbands between the transmissions of different M2M entities. So, UF-OFDM offers better time-frequency efficiency for cognitive M2M system compared to FBMC.

{\bf{\textit{Prior Works.}}}
As of today, literature about resource allocation in CR networks is vast. However, they are either focused on the multicarrier modulation scheme OFDM or FBMC. For instance, in \cite{Bansal}, the authors investigated an optimal power loading algorithm for OFDM-based CR system and studied the impact of subcarrier nulling mechanism on the performance. In \cite{Kang1}, the authors studied the optimal power allocation strategies for achieving ergodic and outage capacity of a CR system under various types of fading models and power constraints. In \cite{Kang2}, the authors proposed a new optimization criterion referred to as \textit{rate loss constraint} instead of conventional interference power constraint to CR system€™s resource allocation optimization. In \cite{Zhang}, the authors considered an OFDM-based CR system and formulated the resource allocation optimization problem as a multidimensional knapsack problem and proposed a greedy \textit{max-min} algorithm to solve it. There is also a plethora of literature about FMBC-based CR networks. In \cite{Haijian}, the authors introduced FBMC as a potential candidate for CR systems and performed investigation on spectral efficiency by balancing the trade-off between interference power caused to the PUs and throughput of SUs. In \cite{Dikmese}, authors studied the effects of combined spectrum sensing and resource allocation for FBMC-based CR system under frequency selective fading channel. In \cite{Yahia}, the impact of time synchronization error on the performance of multicarrier based CR system in a spectrum coexistence context considering CP-OFDM and FBMC modulation schemes was studied. In \cite{Shaat}, the authors proposed a suboptimal solution for the problem of resource allocation in multicarrier-based CR system considering both modulation schemes, OFDM and FBMC, and compared their performance. To the best of the authors' knowledge, no study so far has analysed the performance of UF-OFDM based CR system, especially in the context of cognitive M2M networks.

{\bf{\textit{Rationale and Contributions.}}}
Motivated by the above-mentioned advantages of UF-OFDM over OFDM and FBMC, and to enable low latency, energy-efficient small packet and small signaling transmission in cognitive M2M networks, this paper deals with UF-OFDM based cognitive M2M system. 
 In view of the following key features, this study comes out to be novel and competitive when compared to the existing literature.

\begin{itemize}
\item Two pivotal techniques are properly combined together: {\it{i}}) cognitive M2M and {\it{ii}}) UF-OFDM multicarrier modulation, in order to realize low-latency, small packet machine type communications.  
\item With the aim of building a computationally reasonable resource allocation scheme in order to optimize the capacity of the cognitive M2M system while protecting the primary system, i.e., keeping the interference power introduced to the primary transmission under pre-specified interference threshold, we propose a two-stage solution by dividing the original combinatorial optimization problem into two subproblems. The proposed scheme is found to be computationally efficient.
\item While the subband allocation problem is easily solved via an iterative process, the power allocation problem is handled by transforming it into a second order cone programming (SOCP) optimization problem, which is convex, thus easier to solve.
\item Comprehensive simulation results over realistic propagation scenarios support desired performance characteristics of the cognitive M2M systems.
\end{itemize}

{\bf{\textit{Organization.}}}
The organization of the paper is as follows. While Section \ref{SM} describes the system model and performs instantaneous interference analysis of the UF-OFDM system, Section \ref{PS} provides the problem statement of the resource allocation optimization issue. In Section \ref{PSB}, we propose one low-complexity near-optimal solution. We compare the performance of UF-OFDM based CR system with that of the OFDM- and FBMC-based CR systems in Section \ref{PA}. Finally, Section \ref{CC} concludes the paper.

\section{System Model }
\label{SM}

Let us consider a CR network, where two systems: ({\it{i}}) the primary system and ({\it{ii}}) the secondary downlink cognitive M2M system, are allowed to coexist and operate in the same frequency range. The primary system refers to the one having license to the legacy spectrum. The users belonging to the primary system have the highest priority to access the assigned spectrum. The secondary M2M system refers to the unlicensed cognitive M2M system and, the users/machines belonging to it, referred to as secondary machines (SMs) can only opportunistically access the spectrum holes not used by the primary system.

\begin{figure}
  \centering
   \includegraphics[scale=.055]{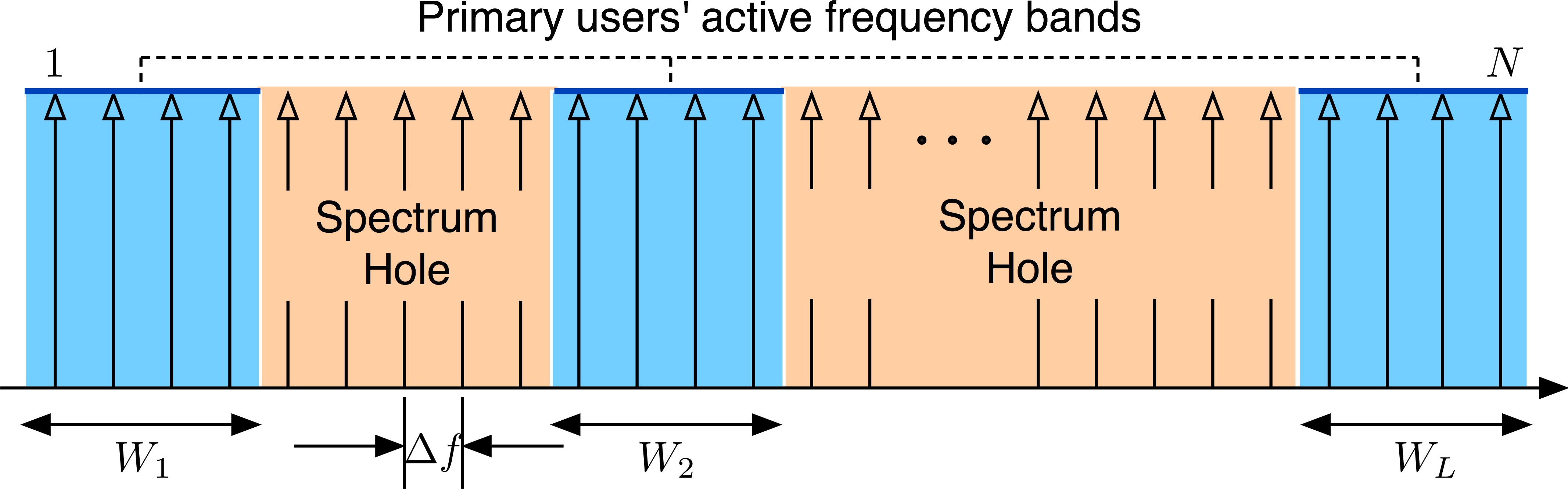}
   \caption{Frequency distribution of the active and nonactive primary bands. Primary users' active frequency bands, spectrum holes and SM OFDM subbands. }
   \label{FRD}
\end{figure}

It is assumed that the CR network consists of $K_{PU}$ PUs who occupy frequency bands of bandwidths, $W_1, W_2, \cdots,W_{K_{PU}}$ Hz, respectively. There are $K_{SM}$ SMs in cognitive M2M system. Let $\mathcal{K}_{PU}$ and $\mathcal{K}_{SM}$ denote the sets of user indices belonging to primary and secondary transmissions, respectively. For simplified analysis, we consider an overlay spectrum sharing approach, i.e., secondary transmission occurs on unoccupied bands only or the SMs are allowed to acquire spectrum resources that are not used by PUs. This approach minimizes the interference to the primary network. The whole bandwidth is divided into $N_{\rm{total}}$ subbands and the width of each subband equals $\Delta f$ Hz. Let  PU $l$ occupy a set of contiguous subbands, $\mathcal{S}_l$ ($W_l$ is integer multiples of $\Delta f$ ) with $N_l=\text{cardinality of }\mathcal{S}_l $. Therefore, the set of subbands available for secondary transmission is given by $\mathcal{N}_{SM}=\mathcal{N}\setminus \mathcal{N}_{PU}=\{f_1,f_2,\cdots,f_{N_{SM}}\}$, where $\mathcal{N}$ is the set of all available subbands, and  $\cup_{l=1}^{K_{PU}}\mathcal{N}_l=\mathcal{N}_{PU}$ is the set of subbands used for primary transmission. The distribution of active and nonactive subbands in the CR system is shown in Fig.~\ref{FRD}.
Therefore, the number of subbands available for secondary transmission equals $N_{SM}=N_{\rm{total}}-( N_l+ N_2+\cdots+ N_{K_{PU}})=\text{cardinality }(\mathcal{N}_{SM})$. $K_{SM}$ SMs share $N_{SM}$ subbands among themselves. We also consider that the modulation format or access mechanism employed by the PUs is unknown to the cognitive M2M system. 

In the downlink transmission of CR network model considered, in general, we have three instantaneous channel fading gains: (i) channel gain between secondary base station (SBS) and SM  $k$ for the $n$th subband denoted as ${\rm{H}}_{kn}^{ss}$; (ii) channel gain between SBS and $l$th PU ${\rm{H}}_{kn}^{sp}$; and (iii) channel fading gain between $l$th PU transmitter and SM  $k$ for the $n$th subband denoted as ${\rm{H}}_{kn}^{ps}$. We consider that these instantaneous channel fading gains are accurately known at SBS. 
However, since PU receiver and SM are not co-located, we assume that SBS can estimate ${\rm{H}}_{kn}^{sp}$ from the transmitted signal of PUs by any existing technique such as hidden feedback approach.

With an ideal coding scheme, the transmission rate achieved by SM $k$ on $n$th subband ($n\in\mathcal{N}_{SM}$), ${\rm{R}}_{kn}$, calculated using Shannon capacity formula, is given by
\begin{equation}
{\rm{R}}_{kn}({\rm{P}}_{kn},{\rm{H}}_{kn}^{ss})=\Delta f\log_{\rm{2}}\left(1+\frac{{\rm{P}}_{kn}|{\rm{H}}_{kn}^{ss}|^2}{\sigma_n^2+\sum_{l=1}^{K_{PU}}{\rm{J}}_{kn}^{(l)}}\right)
\end{equation}
where ${\rm{P}}_{kn}$ is the corresponding transmit power and $ \sigma_{n}^2$ is the mean variance of the additive white Gaussian noise (AWGN) on $n$th subband. $\sum_{l=1}^{{\rm{K_{PU}}}}{\rm{J}}_{n}^{(l)}$ is the amount of interference power introduced into the $n$th subband due to primary transmission on $l$th PU band. We assume that PUs' transmit signals are Gaussian distributed, and SM does not have  knowledge about the codebook employed by the PUs. As a result, we can model the interference introduced to SMs by the PUs as AWGN. However, assuming the interference to be Gaussian distributed may not hold valid for a small number of PU bands, but can be considered a good approximation when a large number of PU bands are available. In what follows, we describe the mathematical models for Interference between the PUs and SMs.

\subsection{UF-OFDM Power spectral Density and Instantaneous Interference Analysis}

UF-OFDM employs Dolph-Chebyshev filters as a feasible ad hoc choice. We have applied this filter (with design parameters $\alpha$ (sidelobe attenuation) and filter order $N$) at the SBS to transmit data to the SMs. Dolph-Chebyshev filter minimizes the Chebyshev norm of the sidelobes for the desired 
main-lobe width, where the filter establishes the sidelobes' norm to -$\alpha$ dB. One notable property of Chebyshev window 
is that sidelobes attenuation remains same at all frequencies, which is unlike the filters employed in OFDM (rectangular window) and FBMC
(PHYDYAS) \cite{Bellanger1} filter that have nonuniform sidelobes attenuation feature.

Based on the $n$th-order Cheyshev polynomials ${{C}_{n}}(\kappa )$ \cite{Lynch}

\begin{equation}
{{C}_{n}}(\kappa )=\left\{ \begin{matrix}
   \cos \left[ n{{\cos }^{-1}}\left( \kappa  \right) \right],\hspace{6mm}\text{for }\left| \kappa  \right|\le 1  \\
   \cosh \left[ n{{\cosh }^{-1}}\left( \kappa  \right) \right],\hspace{2.7mm}\text{for }\left| \kappa  \right|>1 \nonumber  \\
\end{matrix} \right.
\end{equation}
the UF-OFDM filters coefficients in time-domain are defined as
 \begin{equation}
 \begin{array}{*{35}{l}}
\hspace{-3mm}\Psi_{\rm{UF-OFDM}}^{(n)}=\vspace{1.5mm} \\
\vspace{2mm}
\hspace{-3mm}\left\{\begin{matrix} \left\{ \frac{1}{N}+\frac{(10^{-\alpha/20})}{N}2\sum\limits_{m=1}^{M}{{{C}_{2M}}\left[ {{\kappa }_{0}}\cos \left( \frac{\pi m }{N} \right) \right]\cos \left( \frac{2\pi mn}{N} \right)} \right\},\text{for }\left| n \right|\le M  \\
   0,\hspace{65mm}\text{for }\left| n \right|>M  \\
\end{matrix} \right.  \vspace{1.5mm} \\
\end{array}
\end{equation}
where $N=2M+1$ is the UF-OFDM filter length and the parameter ${{\kappa }_{0}}$ is given by ${{\kappa }_{0}}=\cosh \left[ \frac{1}{2M}{{\cosh }^{-1}}\left(10^{\alpha/20} \right) \right]$.

Since UF-OFDM is resource block (RB)-wise filtered, to generate the PDS of any individual subcarrier of UF-OFDM signal, we send zeros at other subband positions of the RB and the UF-OFDM filter response is shifted to the centre frequency of the subband by multiplying the $l$ the coefficient of the filter with $e^{i2\pi(l-1)f_c\frac{1}{N_{\rm{FFT}}}}$. The amplitude of any subband of UF-OFDM signal in time domain is then obtained by performing convolution between OFDM subband and the centre frequency shifted UF-OFDM filter coefficients. Let the amplitude of a single subband in UF-OFDM signal is represented by $\Psi_{\rm{UF-OFDM}}(f )$. Then the power density spectrum (PDS) of a single subband of UF-OFDM is given by
\begin{equation}
\Phi_{\rm{UF-OFDM}}(f )=\left|\Psi_{\rm{UF-OFDM}}(f )\right|^2,
\end{equation}
Since UF-OFDM is RB-wise filtered, outside of the pass-band (subband width) it has a stronger side-lobe decay than OFDM.

\begin{figure}
  \centering
   \includegraphics[scale=.050]{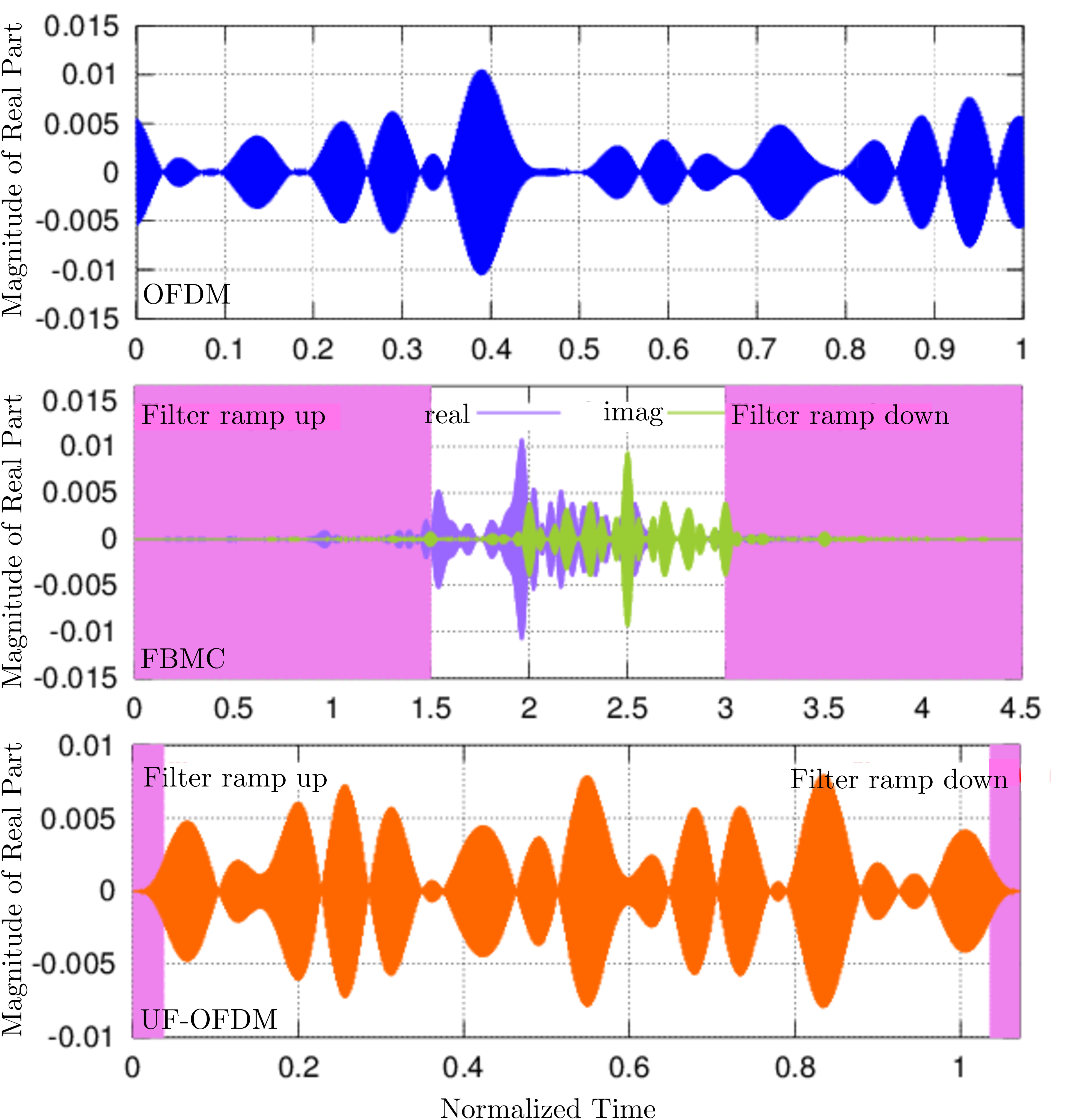}
   \caption{Time domain features of OFDM, FBMC and UF-OFDM waveforms.}
   \label{PC6}
\end{figure}

In Fig.~\ref{PC6}, we show time domain features of OFDM, FBMC and UF-OFDM waveforms. The impulse responses of the prototype filters of the FBMC and UF-OFDM impose transition phases at the beginning (filter ramp up) and at the end (filter ramp down) of the packet as shown with shaded areas. This time domain features are exemplified through packet transmissions in 2 resource blocks (subbands), each subband with 12 subcarriers. For FBMC, the overlapping factor is 4, and for UF-OFDM, the Chebyshev filter length is 74 with side-lobe attenuation of 40 dB. For short packet transmissions as in M2M communications, this long time domain transients are disadvantageous. Compared to FBMC, UF-OFDM offers very short transient time due to its shorter filter length, which makes it more suitable for machine-type communications.

The total interference introduced to any PU band is a result of accumulation of interferences from all the subcarriers available for cognitive M2M transmission.
It is assumed that the cognitive M2M system can use the nonactive PU bands provided that the total interference introduced to the $l$th PU band does not exceed ${I^{(l)}_{\rm{th}}}$, where ${I^{(l)}_{\rm{th}}}$ denotes that the maximum interference power that can be tolerated by the $l$th PU. The interference due to transmission on $n$th subband is given by
 \begin{align}
 {\rm{I}}_{n}^{(l)}({\rm{D}}_{kn}^{(l)},{\rm{P}}_{kn}) & = {\rm{P}}_{kn}\int_{{\rm{D}}_{kn}^{(l)}-{\rm{W}}_{l}/2}^{{\rm{D}}_{kn}^{(l)}+{\rm{W}}_{l}/2}\left|{\rm{H}}_{kn}^{sp}\right|^2\Phi_{\rm{UF-OFDM}}(f)df\nonumber \\ 
 &={\rm{P}}_{kn}\Omega_{kn}^{(l)},
\end{align}
with
 \begin{equation}
 \Omega_{kn}^{(l)}=\int_{{\rm{D}}_{kn}^{(l)}-{\rm{W}}_{l}/2}^{{\rm{D}}_{kn}^{(l)}+{\rm{W}}_{l}/2}\left|{\rm{H}}_{kn}^{sp}\right|^2\Phi_{\rm{UF-OFDM}}(f)df,\nonumber
  \end{equation}
which can be regarded as the interference factor.
Here ${\rm{D}}_{kn}^{(l)}$ quantifies the distance in frequency between the $n$th subband ($n\in\mathcal{N}_{SM}$) and the $l$th PU band. Note that due to the coexistence of primary network and cognitive M2M network in the same frequency range, there exists one more type of interference, which is introduced by the PUs into SMs' bands. For the sake of simplicity, we assume that the interference power introduced to the SMs by the PUs is negligible. Hence, we do not discuss this interference model in details. The interference introduced to the $l$th PU due to secondary downlink M2M transmission is quantified as
\begin{equation}
\sum\limits_{k=1}^{K_{SM}}\sum\limits_{n=1}^{N_{SM}} {\rm{C}}_{kn}{\rm{P}}_{kn}\Omega_{kn}^{(l)}, 
\end{equation}
where ${\rm{C}}_{kn}$ is the indicator function for the subband $n$, i.e., ${\rm{C}}_{kn}=1$ if subband $n$ is assigned to user $k$, and zero otherwise, so that $\sum_{k=1}^{K}{\rm{C}}_{kn}=1,\forall n\in\mathcal{N}_{SM}$, thus meaning that each subband is assigned at most to one user at a time.

\section{Problem Statement }
\label{PS}

 In this UF-OFDM based CR system, our objective is to optimize the total downlink capacity of the secondary cognitive M2M network under total transmit power budget of the SBS while guaranteeing that the interference introduced to the PUs due to secondary transmission remains below the pre-specified interference threshold. Let us define ${\rm{\bf{C}}}=[{\rm{C}}_{kn}]_{K_{SM}\times N_{SM}}$ and ${\rm{\bf{P}}}=[{\rm{P}}_{kn}]_{K_{SM}\times N_{SM}}$. Without loss of generality, we also consider that the pre-specified interference threshold values for all the PUs are the same to avoid the notational clutter. Therefore, ${I^{(l)}_{\rm{th}}}={I_{\rm{th}}}, \forall l, l\in\mathcal{K}_{PU}$. Now the throughput enhancement problem of the cognitive M2M system can be cast as
 
 \begin{equation}
 \label{mainop}
 \begin{array}{*{35}{l}}
\hspace{-3mm}\underset{\{{\rm{\bf{C}}},{\rm{\bf{P}}}\}}{\mathop{\max }}\,\sum\limits_{k=1}^{K_{SM}}{\sum\limits_{n=1}^{N_{SM}}{{{\rm{C}}_{kn}{\rm{R}}_{kn}({\rm{P}}_{kn},{\rm{H}}_{kn}^{ss})}}}\vspace{1.5mm} \\
\vspace{2mm}
\hspace{-3mm}\text{}\text{subject to}  \vspace{1.5mm} \\
\hspace{-3mm}\text{}\hspace{1mm}{{\rm{C}}_{kn}}\in \left\{ 0,1 \right\},\hspace{8mm}\forall k,n,\hspace{1mm}k\in\mathcal{K}_{SM},\hspace{1mm}n\in\mathcal{N}_{SM} \vspace{1.5mm} \\
\hspace{-3mm}\text{}\hspace{1mm} \sum\limits_{k=1}^{{K_{SM}}}{\rm{C}}_{kn}\le1,\hspace{9mm} \forall n,,\hspace{1mm}n\in\mathcal{N}_{SM}\vspace{2mm} \\
\hspace{-3mm}\text{}\hspace{1mm}{{\rm{P}}_{kn}}\ge 0,\hspace{13.5mm}\forall k,n,\hspace{1mm}k\in\mathcal{K}_{SM},\hspace{1mm}n\in\mathcal{N}_{SM} \vspace{2mm} \\
\hspace{-3mm}\text{}\hspace{1mm}\sum\limits_{k=1}^{{K_{SM}}}{\sum\limits_{n=1}^{N_{SM}}{{{\rm{C}}_{kn}}{{\rm{P}}_{kn}}\le {P_{\rm{max}}}}}, \vspace{1.5mm} \\
\hspace{-3mm}\text{}\hspace{1mm}\sum\limits_{k=1}^{K_{SM}}\sum\limits_{n=1}^{N_{SM}} {\rm{C}}_{kn}{\rm{P}}_{kn}\Omega_{kn}^{(l)} \le {I_{\rm{th}}},\hspace{2mm}\forall l\in\mathcal{K}_{PU}\\
\end{array}
\end{equation}
 where ${P_{\rm{max}}}$ is the available transmit power budget of the secondary system. The first two constraints define the non-sharing subband allocation strategy where one subband can be allocated to only one user. The fourth and fifth constraints define the SBS transmit power budget and interference thresholds of the PUs, respectively. This is a combinatorial optimization problem with both integer- and real-valued variables, and accordingly, searching for its global optimum solution is inherently prohibitive. We propose a near-optimal and computationally efficient solution by subdividing the original optimization problem into two subproblems, i.e., {\it{(i)}} subband allocation and {\it{(ii)}} power allocation under given subband allocation. For simplicity and clarity of explanation, we focus only on the case of employing the nonactive subbands by the SMs. The proposed approach can be extended to situations in which both active and nonactive bands are used for cognitive transmission.
 In the following, we describe our proposed near-optimal resource allocation optimization solution to enhance the throughput of the cognitive M2M system.  
 
  \section{Proposed Solution}
  \label{PSB}

For the proposed computationally efficient resource allocation solution, we follow a two-phase optimization process. In the first phase, we employ a suboptimal subband allocation scheme, where the available subbands are allocated to the SMs in such a way all the SMs are treated fairly. In the second phase, for the given subband allocation, we optimally allocate the available SBS transmit power among the SMs.  
 
{\bf{Phase 1:}} \textit{Subband Allocation-}
 In \cite{Jang, Rhee}, the authors have shown that for a multicarrier communication system, in order to maximize the total throughput, each subband needs to be allocated to the user with the best gain on it. However, a user or a group of users may suffer from poor channel gains due to large path loss and fading. Therefore, although allocating the subbands to the users who have the best gains on them will maximize the system capacity, sometimes it may happen that a user or some users do not get assigned with any subband. From the users' perspective, in our proposed iterative subband allocation scheme, we incorporate a notion of fairness among the users, through allowing the user with minimum achieved rate to get assigned a subband in each iteration. The subband allocation process in provided in {\bf{Algorithm 1}}.

  \begin{algorithm}
  \SetAlgoLined
 \textbf{Initialization}: $\mathcal{N}_{SM}=\{f_1,f_2,\cdots,f_{N_{SM}}\}$\;  $R_k=0$, $\mathcal{S}_k=\emptyset, \forall k,\hspace{1mm}k \in {\mathcal{K}_{SM}}$, $P={P}_{\rm{max}}/N_{SM}$;\
  \BlankLine
  \For{$k=1{\rm{~to~}} K_{SM}$}{
    $ f_n = \arg \underset{f_n^*\in\mathcal{N}_{SM}} {\mathrm{max}} ~{\rm{H}}_{kf_n}^{ss}$; ${{S}_{k}}={{S}_{k}}\cup \left\{ f_n \right\}$\;
        ${{R}_{k}}={{R}_{k}}+{{\log }_{2}}(1+P{\rm{H}}_{kf_n}^{ss})$; $\mathcal{N}_{SM}=\mathcal{N}_{SM}\setminus f_n$\;
       }
  \While{$\mathcal{N}_{SM}\ne \emptyset$}{
   $ k = \arg \underset{k^*\in\{1,\cdots,K\}} {\mathrm{min}} ~R_k$; $f_n = \arg \underset{f_n^*\in\mathcal{N}_{SM}} {\mathrm{max}} ~{\rm{H}}_{kf_n}^{ss}$\vspace{1mm}
     \;\vspace{1mm}
       ${\mathcal{S}_k}={\mathcal{S}_i}\cup \left\{ f_n \right\};{\rm{~}}\mathcal{N}_{SM}=\mathcal{N}_{SM}\setminus f_n$\;\vspace{1mm}
              ${{R}_{k}}={{R}_{k}}+{{\log }_{2}}(1+P{\rm{H}}_{kf_n}^{ss})$\;\vspace{2mm}
                              }
                                \BlankLine
      \caption{Subband allocation among the SMs}
   \label{algo2}
\end{algorithm}

At the beginning of the subband allocation process, initialization of all the variables is performed. $\mathcal{N}_{SM}$ is the set of yet unallocated subband indices and $R_k$ keeps track of the capacity for each SM. Thereafter, the second step, i.e., the {\bf{{\it{for loop}}}} allocates to each SM the unallocated subband that has the maximum gain for that SM. Note that an inherent advantage can be gained by the SMs that are able to acquire their best subbands earlier than others, in particular, for the case of two or more SMs having the same subband as their best. However, this bias is negligible when $N_{SM} >> K$ since the probability of such happening will be very low. 
The third step, i.e., the {\bf{{\it{while loop}}}} assigns subbands to each SM according to the greedy policy that the SM that needs a subband most in each iteration gets to acquire the best subband for it. Since we opt to enforce coarse fairness among the SMs such that all the SMs are treated fairly by the resource allocation processes, the need of a SM in each iteration is determined by the SM who has the least capacity achieved so far until there are no more unallocated subbands.
 
 {\bf{Phase 2:}} \textit{SOCP Based Optimal Power Distribution-}
 For the given subband allocation in {\bf{Phase 1}}, we distribute the power optimally among the SMs in order to maximize the throughput. Since the subband allocation is already done, i.e., ${\rm{\bf{C}}}$ is known, the throughput optimization problem is now a function of only one set of variables, which is ${\rm{\bf{P}}}$. Since we employ a non-sharing subband allocation, we can express the channel gains and transmit power variables as the following
 \begin{equation}
 \label{321}
 {\rm{H}}_{n}^{ss}=\sum_{k=1}^{K_{SM}}{\rm{C}}_{kn}{\rm{H}}_{kn}^{ss}/{\sigma_n^2};\hspace{2mm}{\rm{P}}_n=\sum_{k=1}^{K_{SM}}{\rm{C}}_{kn}{\rm{P}}_{kn},
\end{equation}
i.e., we get rid of binary integer variables ${\rm{C}}_{kn}$, and as a result, the power distribution optimization problem can be cast as 
 \begin{equation}
 \begin{array}{*{35}{l}}
\hspace{-24mm}\underset{\{{\rm{\bf{P}}}\}}{\mathop{\max }}\,{\sum\limits_{n=1}^{N_{SM}}{{{\rm{R}}_{n}({\rm{P}}_n,{\rm{H}}_{n}^{ss})}}}\vspace{1.5mm} \\
\vspace{2mm}
\hspace{-17mm}\text{}\text{subject to}  \vspace{1.5mm} \\
\hspace{-17mm}\text{}\hspace{1mm}{{\rm{P}}_{n}}\ge 0,\hspace{13.5mm}\forall n,\hspace{1mm}n\in\mathcal{N}_{SM} \vspace{2mm} \\
\hspace{-17mm}\text{}\hspace{1mm}{\sum\limits_{n=1}^{N_{SM}}{{{\rm{P}}_{n}}\le {P_{\rm{max}}}}}, \vspace{1.5mm} \\
\hspace{-17mm}\text{}\hspace{1mm}\sum\limits_{n=1}^{N_{SM}} P_{n}\Omega_n^l \le {I_{\rm{th}}},\hspace{2mm}\forall l,\hspace{1mm}n\in\mathcal{N}_{SM}.\\
\end{array}
\end{equation}

In the following, we transform the power distribution optimization problem in \eqref{321} as an SOCP problem.
Note that optimization of the new objective function ${\sum\limits_{n=1}^{N_{SM}}{{{\rm{R}}_{n}({\rm{P}}_n,{\rm{H}}_{n}^{ss})}}}=\sum\limits_{n=1}^{N_{SM}}\log_2\left(1+{\rm{P}}_n{\rm{H}}_{n}^{ss}\right)$ and optimization of the function $ \left(\prod\limits_{n=1}^{N_{SM}}\left(1+{\rm{P}}_n{\rm{H}}_{n}^{ss}\right)\right)^{1/N_{SM}}$ will output the same values for the optimization variables ${\rm{P}}_n$. Let us denote $\zeta_n=1+{\rm{P}}_n{\rm{H}}_{n}^{ss}$. Here, $ \left(\prod\limits_{n=1}^{N_{SM}}{{\zeta}_{n}}\right)^{1/N_{SM}}$ defines the geometric mean of optimization variables ${\zeta}_n$, and it is concave. Furthermore, using $ \left(\prod\limits_{n=1}^{N_{SM}}{{\zeta}_{n}}\right)^{1/N_{SM}}$ instead of $\prod\limits_{n=1}^{N_{SM}}{{\zeta}_{n}}$ gives us the flexibility to transform the optimization problem into an SOCP, which is discussed in the following. 
We can eventually transform the objective function ${\mathop{\max }}\,\hspace{1mm} \left(\prod\limits_{n=1}^{N_{SM}}{{\zeta}_{n}}\right)^{1/N_{SM}}\vspace{2mm}$ into one with hyperbolic constraints \cite{Lobo} as discussed below. The example below is for the case of $N_{SM}=4$.

 \begin{equation}
\label{main55}
 \begin{array}{*{35}{l}}
{\mathop{\max }}\,\hspace{1mm} \upsilon_3\vspace{2mm} \\
\text{}\text{subject to} \vspace{2mm} \\
\eta_i=\zeta_i\ge 0, i=1, \cdots, 4 \vspace{2mm} \\
\upsilon_1^2\le \eta_1\eta_2, \hspace{2mm}\upsilon_2^2\le \eta_3\eta_4, \hspace{2mm}\upsilon_3^2\le \upsilon_1\upsilon_2,\vspace{2mm} \\
\upsilon_1\ge 0,\hspace{2mm}\upsilon_2\ge 0,\hspace{2mm}\upsilon_3\ge 0. \vspace{2mm} \\
\end{array}
\end{equation}
In transforming the geometric mean optimization problem above as an SOCP, we have applied the fact that an inequality of the form $y^{2^k}\le q_1q_2\cdots q_{2^k}$ for $t\in \mathbb{R}$, and $q_1\ge 0,\cdots, q_{2^k}\ge 0$ can be cast as $2^{k-1}$ inequalities of the form $\upsilon_i^2\le \eta_1\eta_2$, where all new slack variables that are introduced need to be $\ge 0$. Here, the constraints of the form $\upsilon_i^2\le \eta_1\eta_2$ are hyperbolic constraints.  For hyperbolic equations in the form: ${{\bm{z}}^{2}}\le xy,\text{~~}x\ge 0,\text{~~}y\ge 0$ with $\bm{z}\in\mathbb{R}^{n\times 1}$ and $x,\hspace{1mm}y\in\mathbb{R}$, the equivalent SOCP is given by ${{\bm{z}}^{\mathrm{T}}}{\bm{z}}\le xy,\text{~~}{x}\ge 0,\text{~~}y\ge 0\text{~~}\Leftrightarrow \left\| \begin{matrix}
   2\bm{z}  \\
   x-y  \\
\end{matrix} \right\|_2\le x+y$ \cite{Lobo}.
This hyperbolic transformation makes the optimization problem a convex one as SOCP problems are the types of convex optimization problems. The detailed process of transforming the geometric mean optimization problem into SOCP can be found in \cite{Lobo}.

\section{Performance Analysis}
\label{PA}
A multicarrier system with 3 PUs and 32 subbands is considered, i.e., $K_{PU} = 3$ and $N = 32$. The values of $\Delta f$ and $P_{\rm{max}}$ are assumed to be 0.3125 MHz and 1 watt (unless otherwise specified), respectively. AWGN of variance $\sigma_n^2=10^{-6}$ is considered. Without loss of generality, the interference induced by PUs to the SMs band is assumed to be negligible. The channel power gains ${\rm{H}}_{kn}^{ss}$ and ${\rm{H}}_{kn}^{sp}$ are Rayleigh distributed random variables with mean equal to 1 and considered to be perfectly known at the SBS. The widths of the PU frequency bands $\{W_1,W_2.\cdots, W_L\}$ and the spectrum holes are randomly generated with the total bandwidth used by the primary system being uniformly distributed between the equivalent of 16 to 20 subbands.

\begin{figure}
  \centering
   \includegraphics[scale=.075]{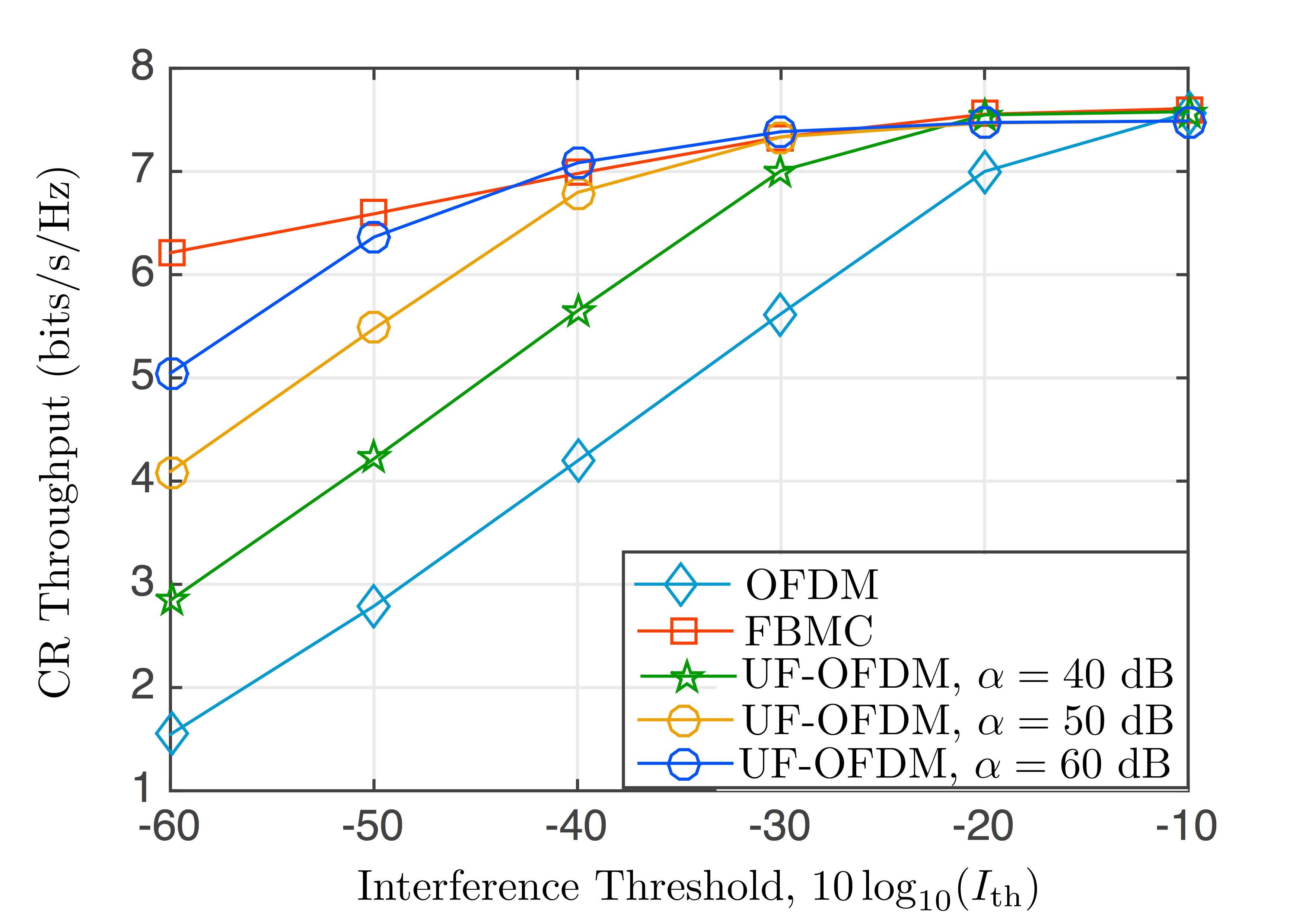}
   \caption{Achieved capacity of the CR system vs. interference threshold tolerated by the PU.}
   \label{PC2}
\end{figure}
The downlink transmission capacity of the SMs versus interference introduced to the PUs' bands is plotted for OFDM, FBMC and UF-OFDM with various sidelobes attenuation factor in Fig.~\ref{PC2}. It is seen that the spectral efficiency of the CR system can be improved by relaxing the interference threshold of the primary system. Furthermore, UF-OFDM modulation based CR system experiences intermediary performance between OFDM and FBMC-based CR systems. OFDM-based CR system achieves the lowest downlink capacity compared to FBMC and UF-OFDM based CR systems. When the maximum amount of interference that can be tolerated by the PUs is very small, i.e., for non-robust PUs, FBMC exhibits the best performance among these three modulation schemes. Note that the achievable downlink capacity for UF-OFDM based system depends on the chosen value for desired sideband attenuation. The higher the value of sideband attenuation factor $\alpha$, the better the achievable downlink capacity. As the interference threshold values increase, therefore, for robust PUs, the achievable capacity of FBMC and UF-OFDM based CR system seem to merge depending on the chosen ripple factor and pre-specified interference threshold. 

\begin{figure}
  \centering
   \includegraphics[scale=.065]{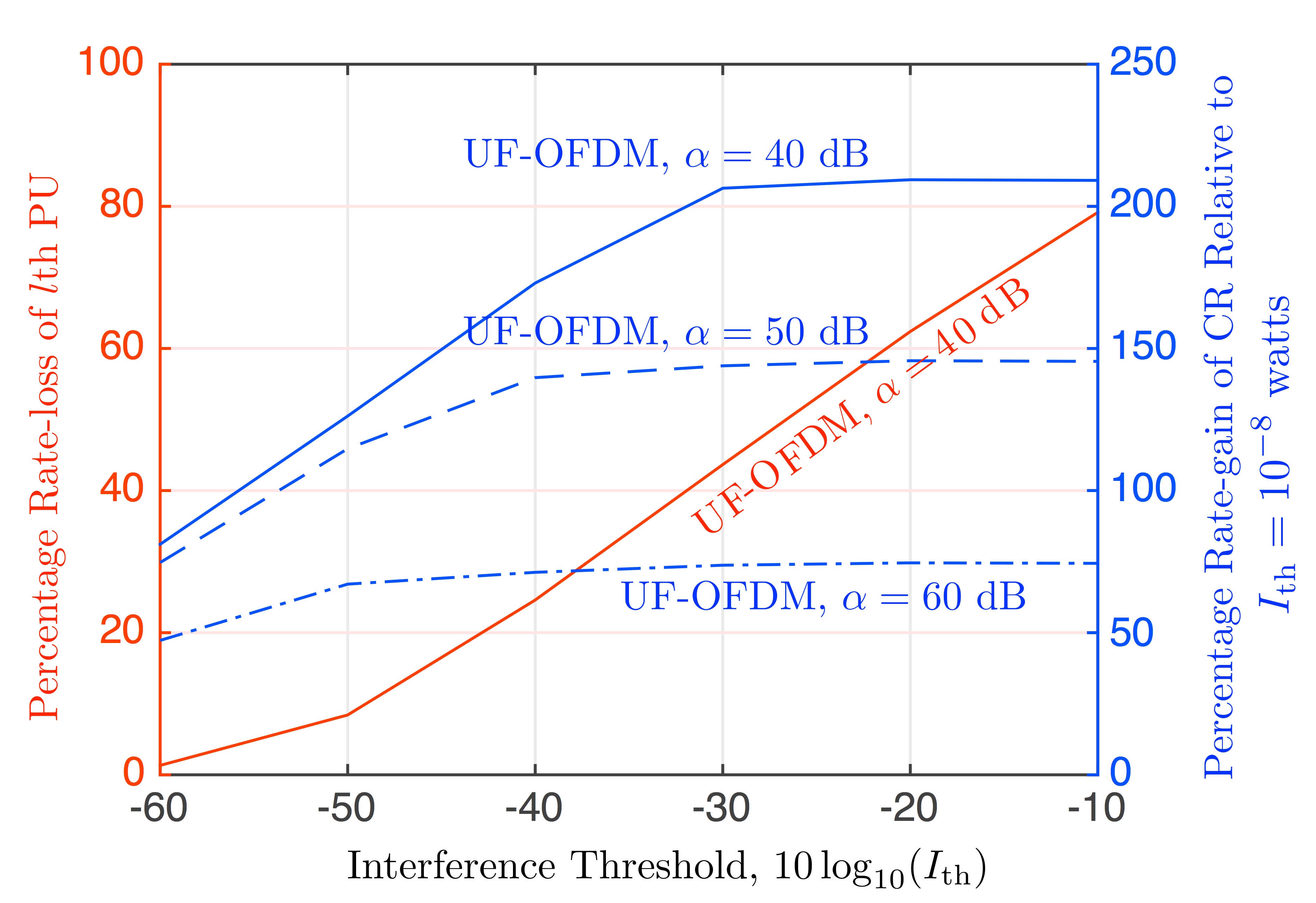}
   \caption{Percentage rate-loss of the $l$th PU versus percentage rate-gain of the PUs as the interference threshold is varied. }
   \label{PC3}
\end{figure}
In Fig.~\ref{PC3}, we show PU's $\%$\textit{rate-loss} versus SMs' $\%$\textit{rate-gain} when pre-specified interference threshold of the PUs is varied. It is observed that the PU's $\%$\textit{rate-loss} increases with $I_{{\rm{th}}}$. For a certain range of $I_{{\rm{th}}}$ values, for example, [-50 dBw -10 dBw], there is a steep rise in the $\%$\textit{rate-loss}. On the other hand, we observe the opposite phenomena for SMs, where the SMs have increase in $\%$\textit{rate-gain} when  $I_{{\rm{th}}}$ increases. The rate of increment of the $\%$\textit{rate-gain} is higher when UF-OFDM with lower sidelobe attenuation factor is employed. As we employ UF-OFDM with higher sidelobes attenuation, the $\%$\textit{rate-gain} decreases by a large amount, almost remains flat, especially when the PUs are very robust.

\begin{figure}
  \centering
   \includegraphics[scale=.086]{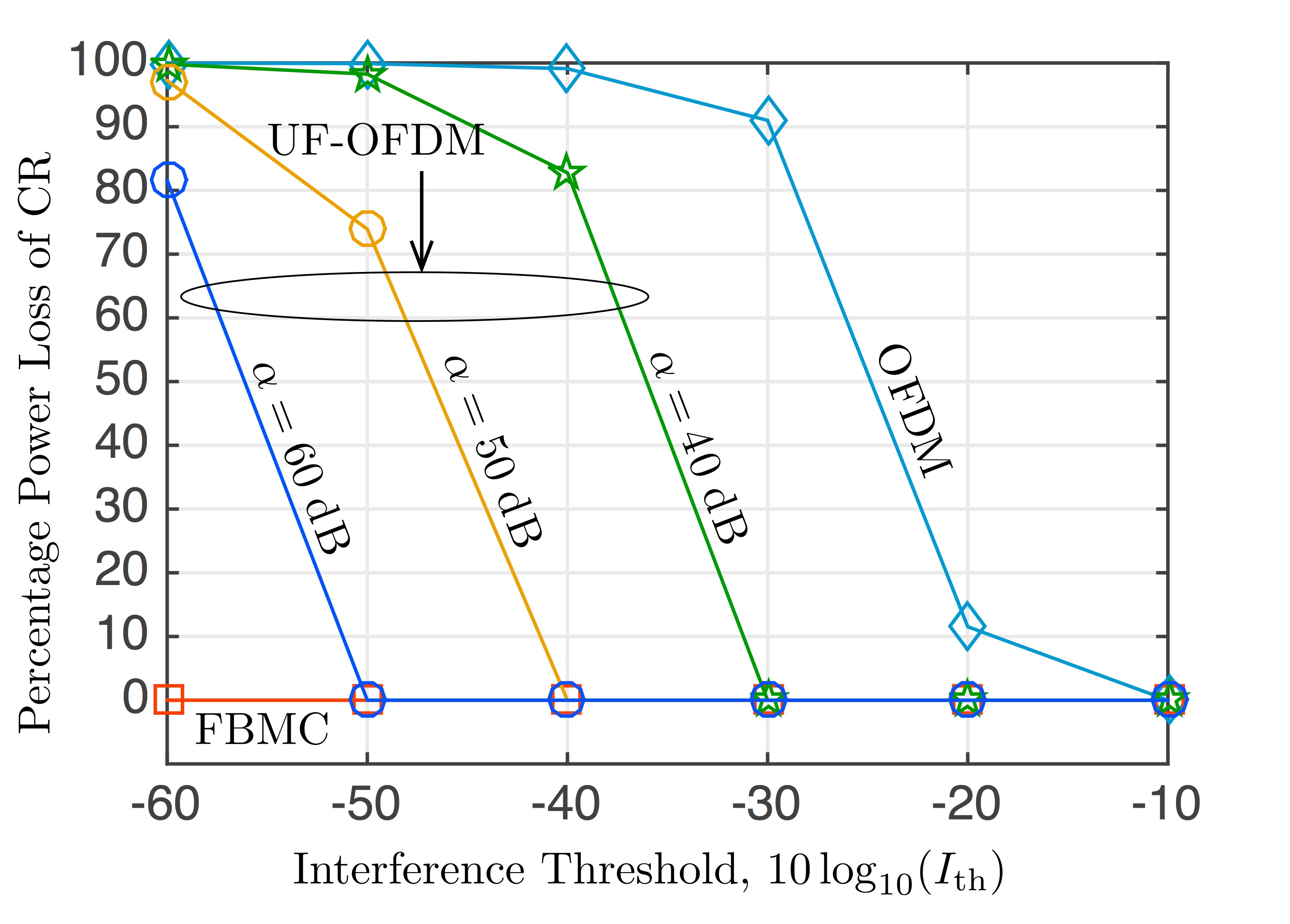}
   \caption{Percentage power-loss of the SBS with $P_{\rm{max}}=1 $ watt. The loss is calculated as, $\%\text{power-loss}=100-P_{{\rm{used}},I_{\rm{th}}}/P_{\rm{max}}\times100$, where $P_{{\rm{used}},I_{\rm{th}}}$ is the actual amount of transmit power used by the SBS in order to satisfy the pre-specified interference threshold $I_{\rm{th}}$.}
   \label{PC7}
\end{figure}
In interference-tolerant CR networks in 5G, a practical and reliable way to manage the mutual interference of CR and primary systems is by regulating the transmit power, which is essential for the CR system to coexist with the primary system. In order to comply with the pre-specified interference threshold of the PUs, the SBS needs to regulate its transmit power, and as a result, sometimes the SBS cannot fully exploit the maximum benefit out of its available power budget. This is clearly demonstrated in Fig.~\ref{PC7}, where we show the $\%$\textit{power-loss} of the SBS for different multicarrier modulation schemes when the pre-specified interference threshold value is varied. It can be observed that OFDM-based CR system experiences the maximum $\%$\textit{power-loss}, whereas FBMC-based CR system does not suffer from $\%$\textit{power-loss} at all over the varying $I_{\rm{th}}$ values. On the other hand, UF-OFDM based system suffers much lower $\%$\textit{power-loss} compared to the OFDM-based CR, and the characteristics of the power-loss curves depend on the robustness of the PUs and predefined sidelobes attenuation factor $\alpha$. For example, when the PUs are moderately robust $(I_{\rm{th}}\in[-30\hspace{1mm} -20])\text{ dBw}$, UF-OFDM-based CR system does not experience any $\%$\textit{power-loss} even with lower $\alpha$. However, for non-robust PUs, the $\%$\textit{power-loss} experienced by the UF-OFDM based CR system increases, and the amount depends on the chosen value of $\alpha$, as can be seen from Fig.~\ref{PC7}.

From the view of computational complexity of the proposed near-optimal resource allocation solution, it is computationally efficient when compared to the original combinatorial optimization problem. For the optimal solution of the combinatorial problem, finding optimal subband assignment requires $K_{SM}^{N_{SM}}$ searches (exhaustive search). Hence, the overall optimization requires $\mathcal{O}(N_{SM}K_{SM}^{N_{SM}})$ operations, which is exponentially complex. Whereas, the computational complexity of our proposed DSP algorithm is composed of two parts, namely (i) subband allocation among the M2M entities in {\bf{Algorithm 1}} with complexity of $\mathcal{O}(K_{SM}N_{SM})$ and (ii) the complexity pertaining to solving SOCP for optimal power allocation, which is polynomial in time\cite{Anderson}. 

\section{Conclusions}
\label{CC}
In this paper, we consider and investigate the resource allocation optimization problem for a UF-OFDM based cognitive M2M network. Instantaneous interference power analysis for UF-OFDM has been carried out. It has been found that UF-OFDM based cognitive M2M performs considerably better than the OFDM-based CR network in terms of achievable cognitive M2M capacity and percentage power loss. For a primary system with robust PUs, the UF-OFDM system works equally well as the FBMC system. However, for the primary system with non-robust PUs, it has been found that using UF-OFDM modulation with higher sidelobe attenuation factor in beneficial. Furthermore, the proposed near-optimal resource allocation solution is computationally very efficient.

Although we have considered that the cognitive M2M system uses only the nonactive or spectrum hole subbands, it is straightforward and can also be demonstrated that the achievable capacity of the cognitive M2M system using the nonactive as well as the active bands is more than that of using the nonactive bands only. Note that in our current study, we have ignored the the interference introduced to the SMs due to primary network's data transmission, which may have noticeable impact on the achievable throughput of cognitive M2M network. We leave consideration of study of impact of interference due to primary transmission on cognitive M2M capacity to future works. Furthermore, perfect channel state information and a small cognitive M2M network have been considered for the sake of simplified performance analysis. We plan to analyse the impact of imperfect channel state information on the achievable capacity of a medium/large cognitive M2M system in our future study.

\section*{Acknowledgment}
\small{This research was conducted under a contract of R\&D for radio resource enhancement, organized by the Ministry of Internal Affairs and Communications, Japan.}

\end{document}